\documentclass[prb,twocolumn,superscriptaddress]{revtex4-1}

\usepackage{amsmath,amssymb,graphicx,color,bm,soul}
\usepackage{textcomp}
\usepackage{hyperref}
\usepackage{color}

\begin{document}
\title{Exciton-Polariton Flows in Cross-Dimensional Junctions}

\author{K. Winkler}
\affiliation{Technische Physik, Wilhelm-Conrad-R\"ontgen-Research Center for Complex
Material Systems, Universit\"at W\"urzburg, Am Hubland, D-97074 W\"urzburg,
Germany}

\author{H. Flayac}
\affiliation{Institute of Physics (iPHYS), \'Ecole Polytechnique F\'ed\'erale de Lausanne (EPFL), Switzerland}

\author{S. Klembt}
\affiliation{Technische Physik, Wilhelm-Conrad-R\"ontgen-Research Center for Complex
Material Systems, Universit\"at W\"urzburg, Am Hubland, D-97074 W\"urzburg,
Germany}

\author{A. Schade}
\affiliation{Technische Physik, Wilhelm-Conrad-R\"ontgen-Research Center for Complex
Material Systems, Universit\"at W\"urzburg, Am Hubland, D-97074 W\"urzburg,
Germany}

\author{D. Nevinskiy}
\affiliation{Department of Electronics and Information Technology, Lviv Polytechnic National University, 12 Bandera street, 79013 Lviv, Ukraine}

\author{M. Kamp}
\affiliation{Technische Physik, Wilhelm-Conrad-R\"ontgen-Research Center for Complex
Material Systems, Universit\"at W\"urzburg, Am Hubland, D-97074 W\"urzburg,
Germany}

\author{C. Schneider}
\affiliation{Technische Physik, Wilhelm-Conrad-R\"ontgen-Research Center for Complex
Material Systems, Universit\"at W\"urzburg, Am Hubland, D-97074 W\"urzburg,
Germany}

\author{S. H\"ofling}
\affiliation{Technische Physik, Wilhelm-Conrad-R\"ontgen-Research Center for Complex
Material Systems, Universit\"at W\"urzburg, Am Hubland, D-97074 W\"urzburg,
Germany}
\affiliation{SUPA, School of Physics and Astronomy, University of St Andrews, St Andrews
KY16 9SS, United Kingdom}

\begin{abstract}
{We study the nonequilibrium exciton-polariton condensation in 1D to 0D and 1D to quasi-2D junctions by means of non-resonant spectroscopy. The shape of our potential landscape allows to probe the resonant transmission of a propagating condensate between a quasi-1D waveguide and cylindrically symmetric states. We observe a distinct mode selection by varying the position of the non-resonant pump laser.
Moreover, we study the the case of propagation from a localized trapped condensate state into a waveguide channel. Here, the choice of the position of the injection laser allows us to tune the output in the waveguide. Our measurements are supported by an accurate Ginzburg-Landau modeling of the system shining light on the underlying mechanisms.}
\end{abstract}

\maketitle{\em Introduction.---} Strong coupling between microcavity photons and quantum well excitons leads to the formation of exciton-polaritons which can in turn form coherent quantum states~\cite{Kasrpzak2006} mediated by bosonic stimulated scattering~\cite{Savvidis2000}. Polariton states are of macroscopic size and can propagate coherently over large distances~\cite{Nelsen2013}\cite{Fischer2014}. Due to their hybrid nature, polaritons in microstructures can be conveniently confined~\cite{Schneider2017} and manipulated~\cite{Wertz2010} via their excitonic or photonic part. Reconfigurable repulsive potentials can be introduced in these systems through a local exciton reservoir generated by a non-resonant laser. For these reasons, polaritons have emerged as a versatile platform to study the behavior of a quantum fluid of light~\cite{Carusotto2013} and as a promising candidate for fast modulated~\cite{Miller2010} optical on-chip logic circuits~\cite{Liew2010} in a solid state environment.
Progresses in this field have yielded logic elements such as an all-optical router~\cite{Flayac2013}\cite{Marsault2015}, a transistor~\cite{Gao2012}\cite{Ballarini2013}, an amplifier~\cite{Wertz2012}\cite{Niemitz2016}, or a Mach-Zehnder interferometer~\cite{Sturm2014}. A precise and deterministic control over the confined polariton modes and polariton propagation into waveguides is essential for an all optical circuit.

In this work, we generate an exciton-polariton condensate in an AlGaAs/AlAs Fabry-P\'erot microcavity via nonresonant injection and investigate the spatial spread at the interface of potentials of different dimensionality. These potentials are composed of cylindrical traps connected to a waveguide wire. In such "lollipop"-shaped potential landscapes, we generate a condensate via nonresonant pumping in the wire section, which expands over the device. Furthermore we demonstrate the possibility to feed into the waveguide wire from the trap while controlling the energy of the injecting mode by spatial positioning of the laser spot.

\begin{figure}[h!]
  \includegraphics[width=0.41\textwidth]{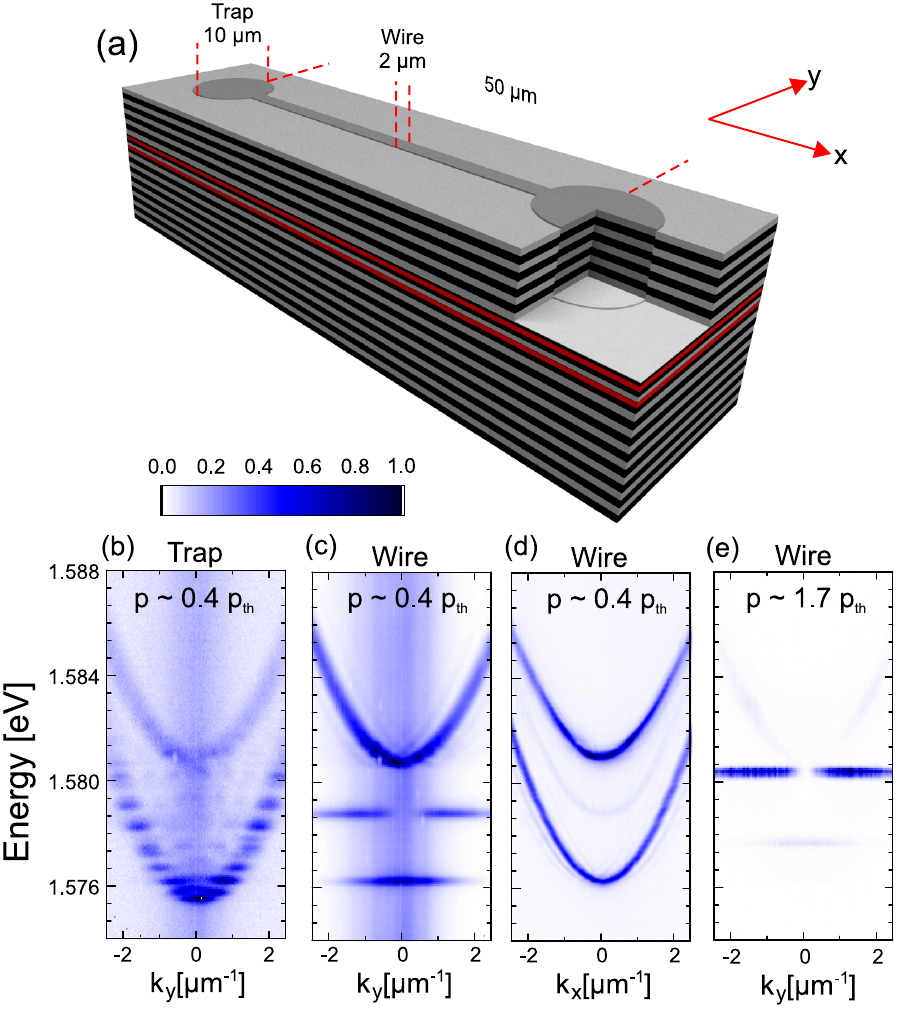}
  \caption{(a) Scheme of the investigated microcavity device featuring a photonic dumbbell potential landscape. The potential of two 10 \textmu m diameter traps connected by a 2 \textmu m wide and 40 \textmu m long wire is introduced due to a locally elongated cavity layer. (b)-(e) Momentum resolved luminescence spectra measured at trap (b) and wire (c)-(e) of the device for excitation power below (b),(d) and above (e) condensation threshold: (a,b,e) measured along $y$-direction for $k_x=0$ whereas (d) measured in $x$-direction for $k_y=0$.
  }\label{Fig:1}
\end{figure}

\textit{Experiment.---} We investigate the spreading of polariton condensates in lollipop-shaped potentials by means of realspace imaging as the polariton wave functions in confined systems are directly observable~\cite{Kalevich2015}\cite{Nardin2010}. The potential itself is generated by a well controlled local elongation of the cavity layer thickness~\cite{Daif2006}. The microcavity grown by molecular beam epitaxy [see Fig.~\ref{Fig:1}(a)] consists of an AlAs-$\lambda/2$-cavity surrounded by 37 (32) bottom (top) AlGaAs/AlAs distributed Bragg reflector (DBR) mirror pairs while the active media comprises two stacks of four $7$ nm GaAs quantum wells distributed at the optical antinodes inside the cavity layer and at the first DBR interface [red color coded in Fig.~\ref{Fig:1}(a)].

By patterning the $10$ nm thick GaAs-layer on top of the cavity layer in an etch-and-overgrowth step, we introduce an attractive photonic potential with a height-difference of about $5$ nm which amounts to about $5$ meV. More details on the process and further sample details concerning mode confinement and condensation properties can be found elsewhere~\cite{Winkler2015}. As exciton-polaritons are hybrid light-matter quasiparticles, the photonic potential yields a trapping potential for polaritons~\cite{Boiko2008}\cite{Kaitouni2006}, which also supports polariton condensation~\cite{Winkler2015}.

Here we carry out measurements of three devices which differ in the in-plane geometry of the potential. Exemplarily a sketch of the 10-\textmu m-dumbbell potential is depicted in Fig.~1(a): It is composed of two circular traps with a diameter of 10 \textmu m connected via a 40 \textmu m long, 2 \textmu m wide input wire. The three devices which are subject to this investigation essentially differ by the diameter of the circular traps (5, 10, and 20 \textmu m) while the wire dimensions are unchanged.

\begin{figure}[htb]
  \includegraphics[width=0.40\textwidth]{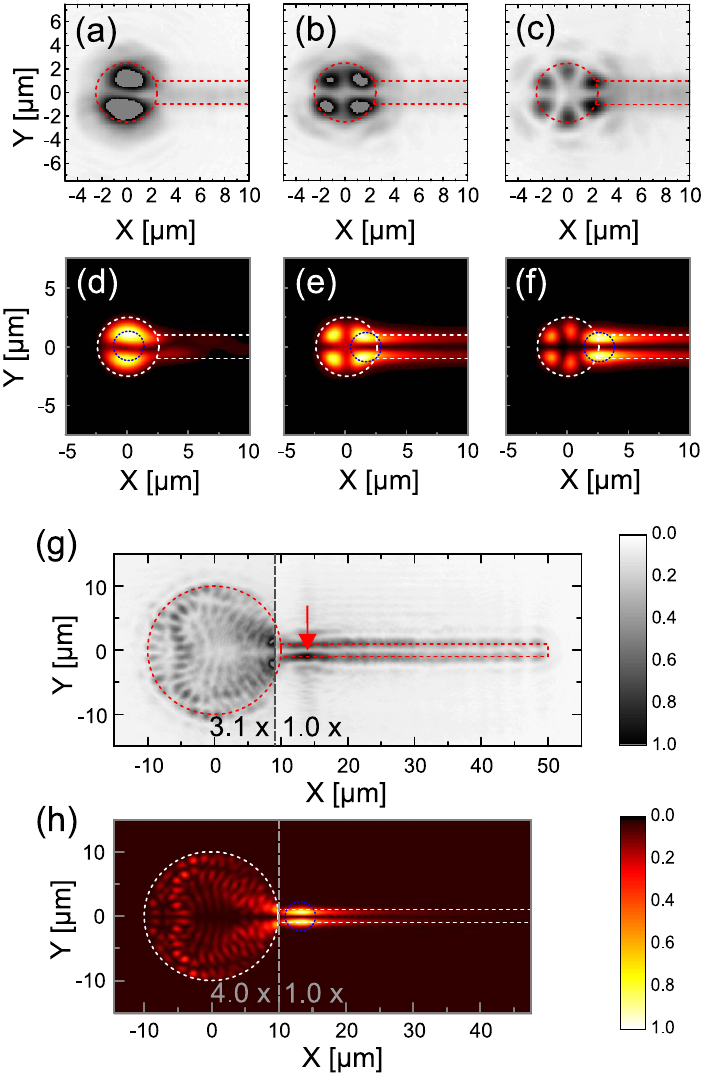}\\
  \caption{Real-space photoluminescence emission (experiment plotted in grayscale, numerical simulations in colorscale throughout the paper) under excitation with a focused Gaussian laser spot. Boundaries of the potential landscapes are indicated by dashed lines.
  (a)-(c) Selection of different orbital modes in the 5-\textmu m-lollipop potential while exciting the trap stepwise at different positions in the $x$-direction.
  Numerically derived real-space images are plotted in (f)-(h), each corresponding to the panel above.
  (g) 20-\textmu m-lollipop excited at the wire giving rise to a heart-shape lobe pattern at the trap which is reproduced by simulation (h).}
 \label{Fig:2}
\end{figure}

\textit{Results and discussion.---}
Figures~\ref{Fig:1}(b)-(e) show angular-resolved energy spectra under continuous wave nonresonant excitation at the 10-\textmu m-dumbbell potential. The transition from the circular traps to the wire is detected at a exciton-photon detuning of $\Delta$ = $E_c$ - $E_x$ = -15.9 meV for the groundstate of the trap to $\Delta$ = -10.1 meV for the free propagating state while the Rabi splitting amounts to 11.5 meV. In this detuning range the fundamental mode polaritonic linewidth extracted from the $k_y=0$ line profile is $248\pm9 ~\mu eV$. As the dimensions of trap and wire differ, the ground mode in the wire is slightly blueshifted with respect to the trap and eigenstates of the system change from 0D cylindrical modes to 1D-wire modes. Several discrete modes are visible in the trap region, together with a continuous parabolic dispersion which stems from the planar background cavity mode. At the position of the wire, two modes can be identified, quantized only in the $y$-direction [see Figs.~\ref{Fig:1}(c) and 1(d)]. With increased injection power, condensation in the wire region takes place in the second mode [see Fig.~\ref{Fig:1}(e)] as witnessed by the characteristic two lobe pattern.

Such a behavior is quite counterintuitive at first sight. Indeed, due to phonon-assisted energy relaxation the condensation in the wire should take place in the lowest energy mode with the single lobe symmetry. However, the slight asymmetry of the sample along the $y$-direction combined with an imperfect positioning of the pump spot makes the condensate to systematically target a combination of the quantized wire modes selected by the overlap between the excitonic reservoir and the eigenmodes of the structure [see appendix~\ref{AppA}]. With increasing pump power and therefore local blueshift the second quantized mode of the wire [see Figs.\ref{Fig:1}(c) and 1(e)] is preferentially selected.

The system dynamics is modeled at the mean field level by means of a 2D driven-dissipative Ginzburg-Landau equation for the polariton wavefunction $\psi(x,y,t)$
\begin{eqnarray}\label{psixyt}
  \left( {i  - \eta } \right)\hbar\frac{{\partial \psi }}{{\partial t}} = &-& \frac{{{\hbar ^2}\Delta }}{{2m}}\psi + U\psi + \alpha {{\left| \psi  \right|}^2}\psi \\
\nonumber   &+& \frac{{i\hbar }}{2}\left( {P - \gamma  - \Gamma {{\left| \psi  \right|}^2}} \right)\psi
\end{eqnarray}
Here $U(x,y)$ is the potential landscape imposed by the structure, $\alpha=10^{-3}$meV$\cdot\mu$m is the polariton scattering strength, $P(x,y)=50\gamma\exp[-(x-x_0)^2/dx^2]\exp[-(y-y_0)^2/dy^2]$, $\gamma=5\times10^{-2}$ ps$^{-1}$ and $\Gamma=0.1\gamma$ are, respectively, the gain from the Gaussian nonresonant pump of extension $dx=dy=2.5\mu$m positioned at $(x_0,y_0)$, polariton loss rate and gain saturation magnitude. Finally $\eta=10^{-2}\gamma$ is a phenomenological energy relaxation rate \cite{Tosi2012} modeling the inelastic phonon scattering characteristic of nonresonant excitation. We note that a background phase and amplitude noise is added in order to allow for excited state stimulation.

We begin our discussion with the smallest structure characterized by a trapping region of 5 \textmu m in diameter. By varying the pump spot position $x_0$ close to the trap, we were able to select the orbital quantum number of the trapped quantized mode as shown in Figs.~\ref{Fig:2} (a)-2(c). The wave functions of these cylindrical modes are described by a radial ($n=1, 2,...$) and orbital ($|m|=0, 1, 2,...$) quantum number where $n$ corresponds to the number of radial nodes and $2\cdot m$ to the number of orbital nodes. The orbitals $m=1,2,3$ of the $n=1$ mode with their characteristic 2, 4, and 6 lobes pattern are excited by gradually imposing $x_0=0$, 1.5, and 2.5$\mu$m, respectively. These modes are enforced by the continuity of the wave function at the boundary with the guide which, as discussed above, favors the two lobe profile which in turn imprints its symmetry to the trap modes. The spot position determines the overlap of the exciton reservoir \cite{Bajoni2008} with the trap modes and selects the angular momentum [see appendix~\ref{AppA}] as confirmed by our simulations [see Figs.\ref{Fig:2}](d)-2(f). Since the selection of a distinct mode comes together with an energy selection we can utilize this to control the polariton condensate flow into the wire.

When excited at the wire, polaritons flowing away from the laser spot act as a resonant injection into high-orbital modes of the trap. This excitation scheme allows for a separation of the background exciton reservoir formed by the pump spot and the emission at the trap by several micrometers. Therefore interactions of the condensate with an incoherent exciton background are reduced~\cite{Schmutzler2014} and thermalization is mostly governed by the scattering rates between 0D states which are dependent on the trap size~\cite{Paraiso2009}.

\begin{figure}[h]
\includegraphics[width=0.46\textwidth]{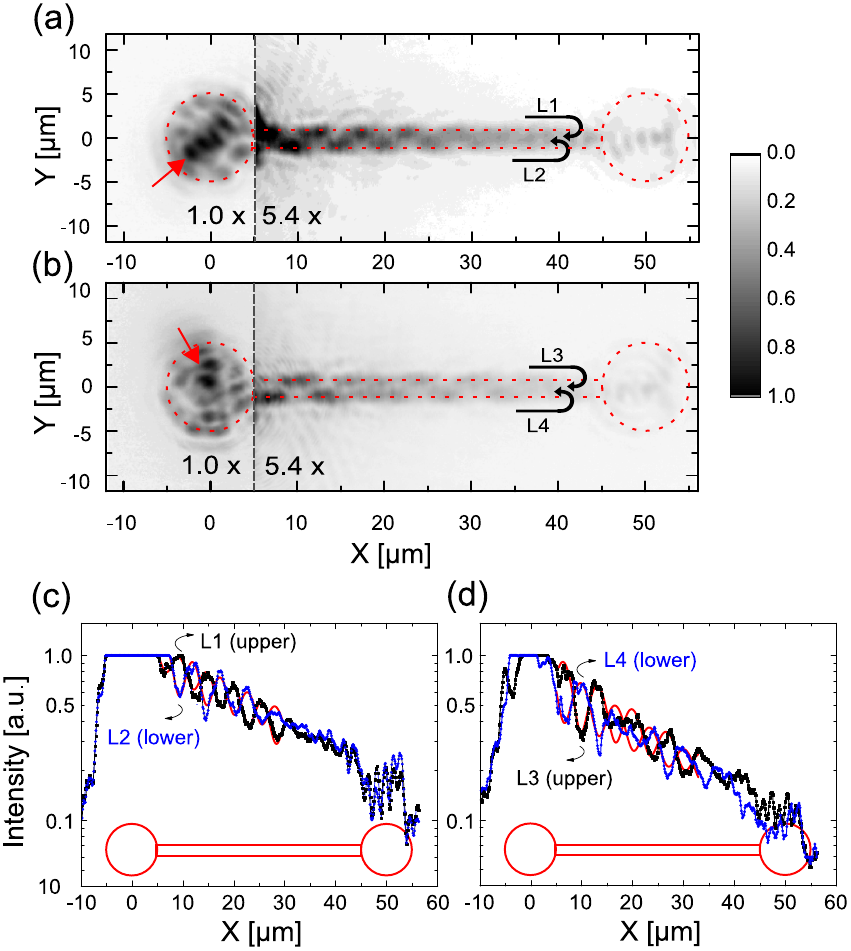}
\caption{(a),(b) Real-space photoluminescence emission under excitation with a focused laser spot (position indicated by red arrow) at the trap of the 10-\textmu m-dumbbell potential. The polariton condensate propagates into the wire and shows an oscillation pattern which is in antiphase for upper (L1/L3) and lower (L2/L4) parts of the wire. By rotating the laserspot position from (a) to (b) the oscillation pattern changes its symmetry. (c),(d) Line profiles along the $x$-direction in detail taken from (a) and (b). Small red arrows indicate a change in the potential landscape.}
\label{Fig:3}
\end{figure}

Now in the case of a 20-\textmu m-wide trap [see Fig.~\ref{Fig:2}(g)] the eigenstates of the trap are quasidegenerated and nearly free propagation is possible. When pumping over the wire, the local blueshift is converted to kinetic energy and particles are expelled in both directions~\cite{Wertz2012}, a stationary scattering pattern, imposed by the wire mode, is visible in the trap. This mode resembles a double heartshape with an entry angle of about 69 degrees which is reproduced by solutions of Eq.~(\ref{psixyt}) [see Fig.~\ref{Fig:2}(h)].

\begin{figure}[htbp]
\includegraphics[width=0.45\textwidth]{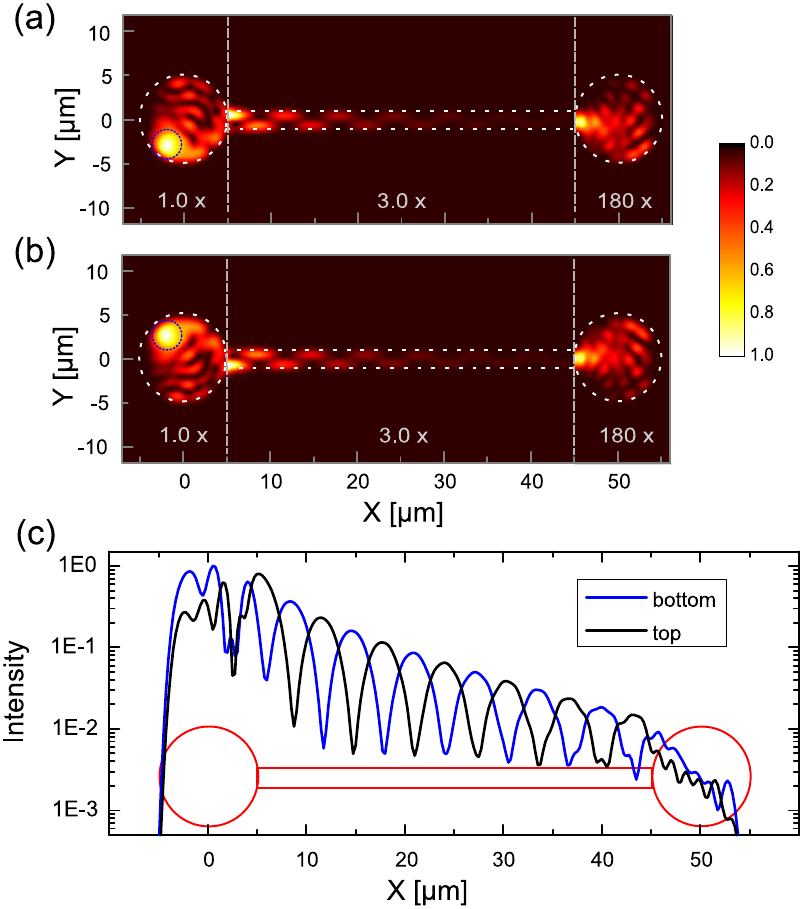}
\caption{(a),(b) Numerical simulations that reproduce the excitation condition in Fig.~3. The position of the Gaussian pump is marked by a dashed blue circle and the potential landscape is indicated by white dashed lines. (c) Line profiles parallel to the wire extracted from (a)}.
\label{Fig:4}
\end{figure}

Our potential landscape furthermore allows us to select the nature of the guided modes by tuning the position of the pump spot on the trap. We excite the trap of the 10-\textmu m-dumbbell at the very edge and a distinct angle to inject into a state of high orbital momentum and thus an energy close to the unbound states [see Figs.~\ref{Fig:3}(a) and 3(b), laser position is marked by a red arrow]. Under such excitation, we excite a combination of both wire modes and therefore the condensate can propagate into the wire. As the n=1 and n=2 state of the wire are excited with different wave vectors, an interference-induced intensity oscillation is visible in the real-space image in the wire. The observed pattern is similar to the one obtained in the context of multimode interferences~\cite{Soldano1995}. Two line profiles parallel to the wire [L1 and L2 in Fig.~\ref{Fig:3}(a)] reveal a damped oscillation which is in antiphase for upper and lower part [see Fig.~\ref{Fig:3}(c)]. Such a behavior is accurately reproduced by our simulations in Fig.~\ref{Fig:4}(a).

The combination of propagating wave vectors can be changed by symmetrically injecting polaritons from the opposite direction in respect to $y=0$ [see Fig.~\ref{Fig:3}(b), Fig.~\ref{Fig:4}(b) for according simulations]. In this case the oscillation has its first maximum in the lower part of the wire (L4) which is revealed in the line profiles of L3 and L4 [see Fig.~\ref{Fig:3}(d), Fig.~\ref{Fig:4}(c) according theory]. We underline that by tuning the input intensity and therefore the local blueshift, one can select higher wave vectors and tune the periodicity of the wire emission pattern [see appendix~\ref{AppB}].

\textit{Conclusions.---}
We have demonstrated control over the spread of an exciton polariton nonequilibrium condensate in potentials which are imprinted into a GaAs-based semiconductor microcavity by an etch and overgrowth technique. These potentials are composed of areas with different in-plane dimensionality yielding different states. Utilizing real-space spectroscopy we directly map these states and investigate the resonant injection between the different parts of the potential. We have shown the possibility to use a wire connected to a cylindrical potential as a feed to resonantly inject a distinct mode in a robust manner. We believe that our results are crucial for the advanced design of polariton integrated circuits which interfaces optical elements of various dimensions.

\textit{Acknowledgements.---}
The W\"urzburg group acknowledges the financial support by the state of Bavaria and the `Deutsche Forschungsgemeinschaft' (DFG) within the project Schn1376-3.1. S.K. acknowledges the European Commission for the H2020 Marie
Sk\l odowska-Curie Actions (MSCA) fellowship (Topopolis). We would like to thank I. G. Savenko and M. Sun for discussions.

\
\appendix
\
\setcounter{figure}{0}

\begin{figure}[htbp]
\renewcommand{\figurename}{Fig. A}
\includegraphics[width=0.45\textwidth]{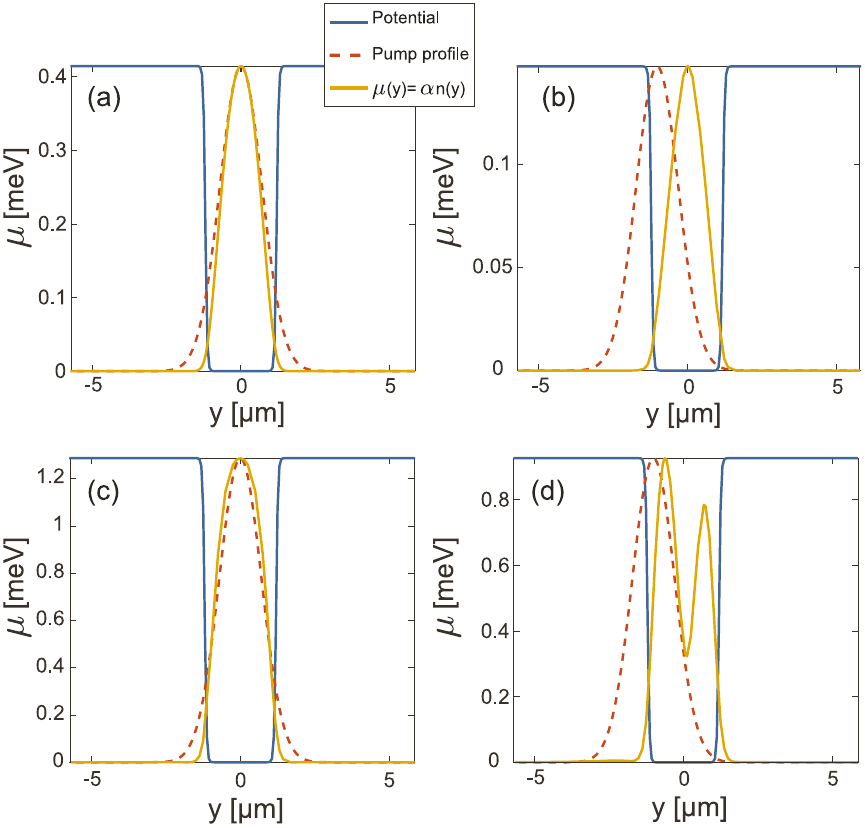}
\caption{Density profile (yellow line) expressed in terms of the local blueshift $\mu(y)=\alpha n(y)$. (a) Centered spot $y_P=0$ (dashed-red line) and pump amplitude $A_P=20~\hbar\gamma$. (b) Shifted spot $y_P=-1$ $\mu$m and pump amplitude $A_P=20~\hbar\gamma$. (c) Centered spot $y_P=0$ and pump amplitude $A_P=30~\hbar\gamma$. (d) Shifted spot $y_P=-1$ $\mu$m and pump amplitude $A_P=30~\hbar\gamma$. The potential profile (blue line) is normalized here to match the maximum of the density profile for the sake of clear visualization.}
\label{Fig:A1}
\end{figure}

\section{Impact of pump shift and power}
\label{AppA}
To shed light on the two lobe pattern obtained by pumping the wire, we resort to an effective 1D theory across the wire. We solve Eq.~\eqref{psixyt} along the $y$ direction and assume full translational invariance in the $x$ direction. We show in Fig.A1 the condensate density pattern obtained for two different pump power and pump spot positions. As one can see below a threshold chemical potential, the condensation occurs in the $n=1$ mode of the wire demonstrating a one lobe pattern. The two lobe $n=2$ pattern emerges from the combination of a sufficient blueshift associated with a sufficient shift of the pump spot as one can see in panel (d) [see captions].

\section{Wire mode superposition}
\label{AppB}
As shown in Fig.~\ref{Fig:3}, the position of the pump spot in the trap allows exciting a combination of the $n=1$ and $n=2$ modes of the wire which is easily reproduced by considering two plane waves quantized along the $y$-direction
\begin{eqnarray}\label{PW}
  {u_1}\left( {x,y} \right) &=& {A_1}\sin \left( {\frac{{{n_1}\pi y }}{{{L_y}}}} \right){e^{i{k_1}x}} \hfill \\
  {u_2}\left( {x,y} \right) &=& {A_2}\sin \left( {\frac{{{n_2}\pi y}}{{{L_y}}}} \right){e^{i{k_2}x}}.
\end{eqnarray}
The resulting intensity pattern $I(x,y)=|u_1(x,y)+u_2(x,y)|^2$ is shown in Fig.A2 in the case $n_1=1$, $n_2=2$, $A_1=1$, $A_2=1$, and $\Delta k=k_1-k_2=2/\mu$m and reproduces the pattern observed in our experiment. The periodicity of the pattern is adjusted via $\Delta k$.

\begin{figure}[htbp]
\renewcommand{\figurename}{Fig. A}
\includegraphics[width=0.45\textwidth]{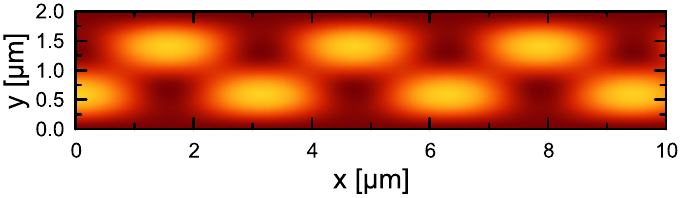}
\caption{Wire interference pattern obtained by superposition of two confined plane waves with different wave vector and transverse quantum number.}
\label{Fig:A2}
\end{figure}

To be more quantitative, we have analyzed with our full 2D model the impact of the spot position in the 10$\mu$m trap along the $y$-direction on the interference pattern. The results are shown in Fig.A3 showing that the periodicity of the pattern can be tuned by varying $y_P$ and $A_P$.
While the interferences are suppressed for a fully centered spot $y_P=0$ in the low density [see panel (a)] and high density regime [see panel (b)] the situation changes as the spot is shifted in the $y$-direction. From power dependent simulations shown in Figs.~A3(d)-(f) for a spot shifted to $y_P=-1~\mu m$ it can be seen that oscillations are suppressed [see panel (d)] until the local blueshift reaches the energy of the $n=2$ wire mode [see panel (e)]. A second mode contribution is visible in the corresponding wire dispersion along $x$ [see panel (g)] in opposition to the centered spot [see panel (c)]. The pump power impacts as it allows targeting higher initial energy and therefore larger pairs of wave vectors, which manifests in a larger periodicity of the oscillation [see panels (e) and (f)].
The attack angles of the waves repelled by the high density pump spot allow us to tune the $\Delta {k_x} = \left( {{{\mathbf{k}}_1} - {{\mathbf{k}}_2}} \right) \cdot {{\mathbf{u}}_x}$ value and therefore the pattern [see panel (h)].

\begin{figure}[h]
\renewcommand{\figurename}{FIG.A}
\includegraphics[width=0.45\textwidth]{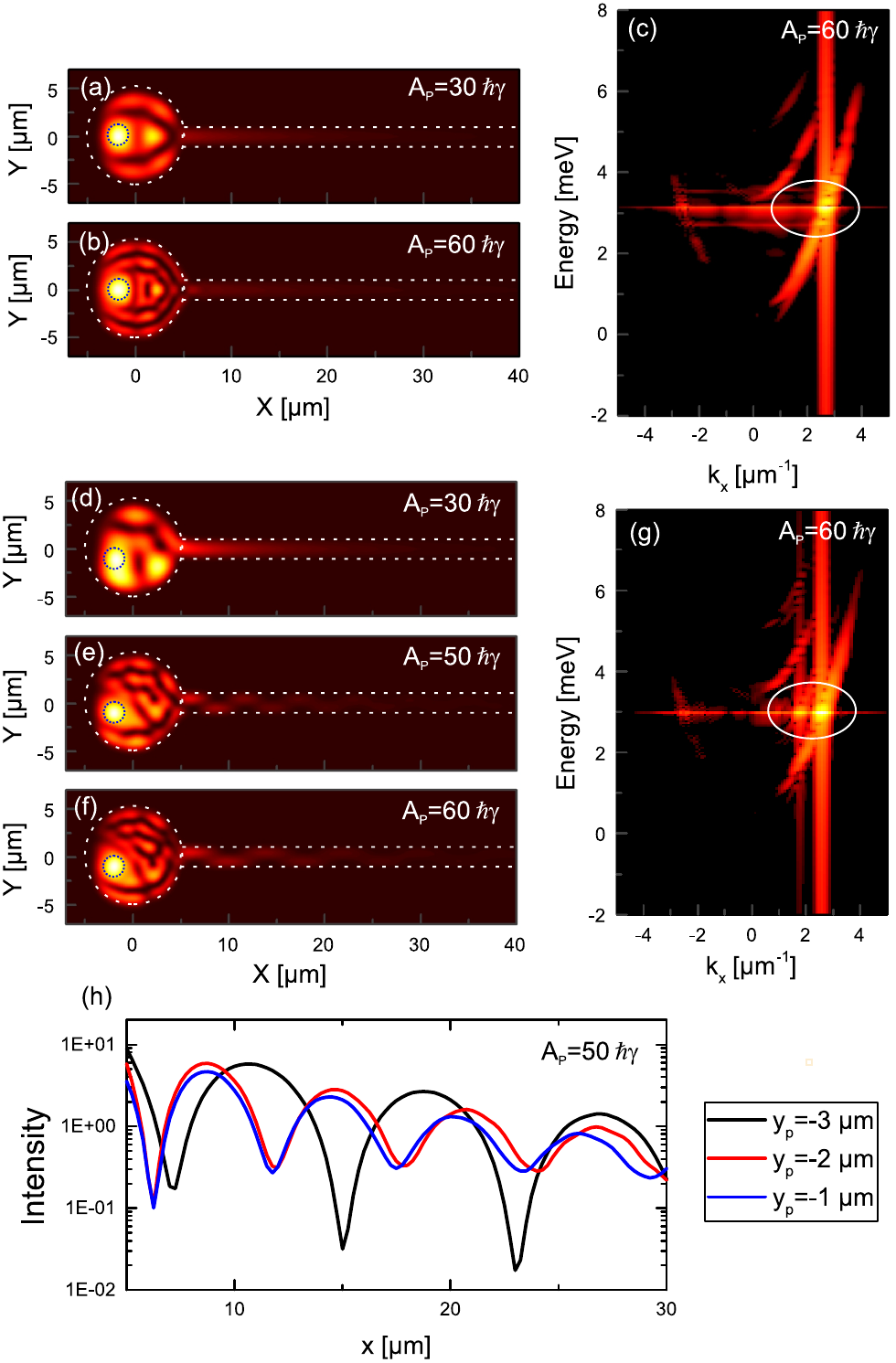}
\caption{Simulation of realspace emission and corresponding dispersions along the wire for different pump amplitudes $A_P$. Spot position is marked by a blue dashed circle and dispersions are plotted in logarithmic scale. (a-c) Excitation by a shifted spot at position $y_P=-1~\mu m, x_P=-2~\mu m$, and corresponding high power dispersion in (d). (e),(f) Excitation by an unshifted spot at position $y_P=0, x_P=-2~\mu m$ together with the dispersion (g) corresponding to (f). (h) Oscillation pattern taken from $y=0.5~\mu m$ and $A_P=50~\hbar \gamma$ for different spot positions $y_P$ while $x_P=-2~\mu m$.}
\label{Fig:A3}
\end{figure}


\clearpage


\begin{thebibliography}{99}

\bibitem{Kasrpzak2006}
J. Kasprzak, M. Richard, S. Kundermann, A. Baas, P. Jeambrun, J. M. J. Keeling, F. M. Marchetti, M. H. Szyman\'{n}ska, R. Andr\'{e}, J. L. Staehli, V. Savona, P. B. Littlewood, B. Deveaud, and Le Si Dang, Nature, {\bf 443}, 409 (2006).

\bibitem{Savvidis2000} P. G. Savvidis, J. J. Baumberg, R. M. Stevenson, M. S. Skolnick, D. M. Whittaker, and J. S. Roberts, Phys. Rev. Lett. {\bf 84}, 1547 (2000).

\bibitem{Nelsen2013} B. Nelsen, G. Liu, M. Steger, D.W. Snoke, R. Balili, K. West, and L. Pfeiffer, Phys. Rev. X {\bf 3}, 041015 (2013).

\bibitem{Fischer2014} J. Fischer, I.G. Savenko, M.D. Fraser, S. Holzinger, S. Brodbeck, M. Kamp, I.A. Shelykh, C. Schneider, and S. H\"ofling, Phys. Rev. Lett. {\bf 113}, 203902 (2014).

\bibitem{Schneider2017} C. Schneider, K. Winkler, M. D. Fraser, M. Kamp, Y. Yamamoto, E. A. Ostrovskaya, and S. H\"ofling, Reports Prog. Phys. {\bf 80}, 16503 (2017).

\bibitem{Wertz2010} L. Ferrier, E. Wertz, R. Johne, D. D. Solnyshkov, P. Senellart, I. Sagnes, A. Lema\^{i}tre, G. Malpuech, and J. Bloch, Phys. Rev. Lett. {\bf 106}, 126401 (2011).

\bibitem{Carusotto2013} I. Carusotto and C. Ciuti, Rev. Mod. Phys. {\bf 85}, 299 (2013).

\bibitem{Miller2010} D.A.B. Miller, Nat. Photonics {\bf 4}, 3 (2010).

\bibitem{Liew2010} T.C.H. Liew, A. V. Kavokin, T. Ostatnick\'{y}, M.A. Kaliteevski, I.A. Shelykh, and R.A. Abram, Phys. Rev. B {\bf 82}, 033302 (2010).

\bibitem{Flayac2013} H. Flayac and I.G. Savenko, Appl. Phys. Lett. {\bf 103}, 201105 (2013).

\bibitem{Marsault2015} F. Marsault, H.S. Nguyen, D. Tanese, A. Lema\^{i}tre, E. Galopin, I. Sagnes, A. Amo, and J. Bloch, Appl. Phys. Lett. {\bf 107}, 201115 (2015).

\bibitem{Gao2012} T. Gao, P.S. Eldridge, T.C.H. Liew, S.I. Tsintzos, G. Stavrinidis, G. Deligeorgis, Z. Hatzopoulos, and P.G. Savvidis, Phys. Rev. B {\bf 85}, 235102 (2012).

\bibitem{Ballarini2013} D. Ballarini, M. De Giorgi, E. Cancellieri, R. Houdr\'{e}, E. Giacobino, R. Cingolani, A. Bramati, G. Gigli, and D. Sanvitto, Nat. Commun. {\bf 4}, 1778 (2013).

\bibitem{Wertz2012} E. Wertz, A. Amo, D.D. Solnyshkov, L. Ferrier, T.C.H. Liew, D. Sanvitto, P. Senellart, I. Sagnes, A. Lema\^{i}tre, A. V. Kavokin, G. Malpuech, and J. Bloch, Phys. Rev. Lett. {\bf 109}, 216404 (2012).

\bibitem{Niemitz2016} D. Niemietz, J. Schmutzler, P. Lewandowski, K. Winkler, M. A{\ss}mann, S. Schumacher, S. Brodbeck, M. Kamp, C. Schneider, S. H\"ofling, and M. Bayer, Phys. Rev. B {\bf 93}, 235301 (2016).

\bibitem{Sturm2014} C. Sturm, D. Tanese, H.S. Nguyen, H. Flayac, E. Galopin, A. Lema\^{i}tre, I. Sagnes, D. Solnyshkov, A. Amo, G. Malpuech, and J. Bloch, Nat. Commun. {\bf 5}, 3278 (2014).

\bibitem{Kalevich2015} V. K. Kalevich, M. M. Afanasiev, V. A. Lukoshkin, D. D. Solnyshkov, G. Malpuech, K. V. Kavokin, S. I. Tsintzos, Z. Hatzopoulos, P. G. Savvidis, and A. V. Kavokin, Phys. Rev. B {\bf 91}, 045305 (2015).

\bibitem{Nardin2010} G. Nardin, Y. L\'{e}ger, B. Pietka, F. Morier-Genoud, and B. Deveaud-Pl\'{e}dran, Phys. Rev. B {\bf 82}, 045304 (2010).

\bibitem{Daif2006}
O. El Da\"{i}f, A. Baas, T. Guillet, J.-P. Brantut, R.I. Kaitouni, J.L. Staehli, F. Morier-Genoud, and B. Deveaud, Appl. Phys. Lett. {\bf 88}, 061105 (2006).

\bibitem{Winkler2015}
K. Winkler, J. Fischer, A. Schade, M. Amthor, R. Dall, J. Ge{\ss}ler, M. Emmerling, E.A. Ostrovskaya, M. Kamp, C. Schneider, and S. H\"ofling, New J. Phys. {\bf 17}, 023001 (2015).

\bibitem{Boiko2008} D. L. Boiko, PIERS Online, {\bf 4}, 831 (2008).

\bibitem{Kaitouni2006}
R. I. Kaitouni, O. El Da\"{i}f, A. Baas, M. Richard, T. Para\"{i}so, P. Lugan, T. Guillet, F. Morier-Genoud, J. D. Gani\`{e}re, J. L. Staehli, V. Savona, and B. Deveaud, Phys. Rev. B {\bf 74}, 155311 (2006).

\bibitem{Tosi2012} G. Tosi, G. Christmann, N.G. Berloff, P. Tsotsis, T. Gao, Z. Hatzopoulos, P.G. Savvidis, and J.J. Baumberg, Nature Communications {\bf 3}, 1243 (2012)

\bibitem{Bajoni2008} D. Bajoni, P. Senellart, E. Wertz, I. Sagnes, A. Miard, A. Lema\^{i}tre, and J. Bloch, Phys. Rev. Lett. {\bf 100}, 047401 (2008).

\bibitem{Schmutzler2014} J. Schmutzler, T. Kazimierczuk, \"O. Bayraktar, M. A{\ss}mann, M. Bayer, S. Brodbeck, M. Kamp, C. Schneider, and S. H\"ofling, Phys. Rev. B {\bf 89}, 115119 (2014).

\bibitem{Paraiso2009} T.K. Para\"{i}so, D. Sarchi, G. Nardin, R. Cerna, Y. Leger, B. Pietka, M. Richard, O. El Da\"{i}f, F. Morier-Genoud, V. Savona, and B. Deveaud-Pl\'{e}dran, Phys. Rev. B {\bf 79}, 045319 (2009).

\bibitem{Soldano1995} L. B. Soldano and E. C. M. Pennings, J. Light. Technol. {\bf 13}, 615 (1995).
\end{thebibliography}
\end{document}